\documentclass[preprint,aps,prd,showpacs,superscriptaddress,nofootinbib]{revtex4}
\usepackage{graphicx,epsfig}
\usepackage{amsmath}
\usepackage{bm}
\usepackage{color}
\usepackage{relsize}
\RequirePackage{xspace}
\begin{document}

\title{\mbox{}\\[10pt]
Order-$v^4$ Relativistic Corrections to $\Upsilon$
Inclusive Decay into a Charm pair}

\author{Wen-Long Sang}
\affiliation{Institute of High Energy Physics, Chinese Academy
of Sciences, \\Beijing 100049, People's Republic of China}
\author{Hai-Ting Chen}
\author{Yu-Qi Chen}
\affiliation{Key Laboratory of Frontiers in
Theoretical Physics, \\The Institute of Theoretical Physics, Chinese
Academy of Sciences, Beijing 100190, People's Republic of China\vspace{0.2cm}}

\begin{abstract}
In this work, we determine the short-distance coefficients for
$\Upsilon$ inclusive decay into a charm pair through relative order
$v^4$ within the framework of NRQCD factorization formula. The
short-distance coefficient of the order-$v^4$
color-singlet NRQCD matrix element is obtained through matching the
decay rate of $b\bar{b}(^3S^{[1]}_1)\to c\bar{c}gg$ in full QCD
to that in NRQCD. The double and single IR divergences
appearing in the decay rate are exactly canceled through the
next-to-next-to-leading-order renormalization of the operator
${\mathcal O}(^3S^{[8]}_1)$ and the next-to-leading-order
renormalization of the operators ${\mathcal O}(^3P^{[8]}_J)$. To
investigate the convergence of the relativistic expansion
arising from the color-singlet contributions, we study the
ratios of the order-$v^2$ and -$v^4$ color-singlet short-distance
coefficients to the leading-order one. Our results indicate that
though the order-$v^4$ color-singlet short-distance coefficient is
quite large, the relativistic expansion for the color-singlet
contributions in the process $\Upsilon\to c\bar{c}+X$ is well
convergent due to a small value of $v$. In addition, we
extrapolate the value of the mass ratio of the charm quark to the
bottom quark, and find the relativistic corrections rise quickly
with increase of the mass ratio.
\end{abstract}


\maketitle
\newpage

\section{Introduction}
\label{introduction}
NRQCD factorization formula~\cite{Bodwin:1994jh} provides a
systematical approach to express the quarkonium inclusive decay rate
(cross section) as the sum of product of short-distance coefficients
and the NRQCD matrix elements.
The short-distance coefficients can be expanded as perturbation series in coupling
constant $\alpha_s$ at the scale of the heavy quark mass $m$. The long-distance matrix elements can
be expressed in a definite way with the typical relative velocity $v$ of the heavy quark in the
quarkonium.

The relativistic corrections to the quarkonium decay and production
have been widely studied. In some processes, the
next-to-leading-order (NLO) relativistic corrections are sizable,
some of which even surpass the leading-order (LO) contributions. For
the quarkonium production, typical examples are the double charmonia
$J/\psi+\eta_c$ production at the B
factories~\cite{Braaten:2002fi,He:2007te} and the double $\eta_c$
production through $\eta_b$ decay~\cite{Sang:2011fw}. For the
quarkonium annihilate decay, it happens in the $J/\psi$
inclusive decay and $J/\psi(\Upsilon)$ inclusive decay into a lepton
pair or charm pair~\cite{Keung:1982jb,Chen:2011ph,Chen:2012zzg}. It
may arouse worries about the relativistic expansion in the NRQCD
approach. In Refs.~\cite{Bodwin:2006ke,Bodwin:2007ga}, the
authors resumed a collect of relativistic corrections to the process
$e^+e^-\to J/\psi+\eta_c$ and found the relativistic expansion is
of well convergence. Similarly, the authors of
Ref.~\cite{Sang:2011fw} studied the process $\eta_b\to
\eta_c(mS)\eta_c(nS)$ and found the relativistic expansion is
convergent very well, though the order-$v^2$ corrections are both
negative and large. In Ref.~\cite{Bodwin:2002hg}, the authors
considered the order-$v^4$ corrections to $J/\psi$ inclusive decay
and found the contribution from the color-singlet matrix element in
this order is not as large as the LO and NLO contributions. The
order-$v^6$ corrections to this process were even calculated in
Ref.~\cite{Chung:TBD}, where the convergence is furthermore
confirmed.

It has been shown that the order-$v^2$ corrections to the process
$\Upsilon\to c\bar{c}+X$ are extremely huge~\cite{Chen:2011ph}. The
ratio of the short-distance coefficient of the order-$v^2$ matrix
element to that of the LO one approaches to $-12$. It
seriously spoils the relativistic expansion. So it urges us to
calculate the order-$v^4$ corrections and investigate the
convergence of the relativistic expansion for this process.
Moreover, since the momenta of some gluons can be simultaneously
soft, there exist complicated IR divergences in the calculations.
It is technically challenging to cancel the IR divergences and
obtain the IR-independent short-distance coefficients through the
color-octet mechanism~\footnote{When we were doing this work, a work
about order-$v^4$ corrections to gluon fragmentation appeared at
arXiv recently~\cite{Bodwin:2012xc}. In that work, the authors
applied the similar techniques to cancel IR divergences.}, and
it also provides an example to examine the NRQCD factorization
formula.

The remainder of this paper is organized as follows. In Sec.~\ref{sec:NRQCD-factorization},
we describe the NRQCD factorization formula for the $\Upsilon$ inclusive decay into a charm pair.
In Sec.~\ref{sec:definitions}, we list our definitions and present the techniques used to
compute the decay rates.
We elaborate the calculations on determining the short-distance coefficients at relative order $v^4$ in
Sec.~\ref{sec:matching}. The QCD corrections to the color-octet operators are also calculated
in this section. Sec.~\ref{sec:discussion} is devoted to discussions and summaries.

\section{NRQCD factorization formula for $\Upsilon\to c\bar{c}+X$}
\label{sec:NRQCD-factorization}

According to the NRQCD factorization formula, through relative order
$v^4$, the differential decay rate for $\Upsilon$
inclusive decay into a
charm pair can be expressed as~\cite{Bodwin:1994jh}:
\begin{eqnarray}
\label{factorization-formula}
d\Gamma[\Upsilon\to c\bar{c}+X]&=&
\frac{dF_1({}^3S^{[1]}_1)}{m^2}\langle \mathcal{O}({}^3S^{[1]}_1)\rangle_\Upsilon
+\frac{dF_2({}^3S^{[1]}_1)}{m^4}\langle \mathcal{P}({}^3S^{[1]}_1)\rangle_\Upsilon
+\frac{dF({}^1S^{[8]}_0)}{m^2}\langle \mathcal{O}({}^1S^{[8]}_0)\rangle_\Upsilon
\nonumber\\
&&
+\frac{dF_3({}^3S^{[1]}_1)}{m^6}
\langle \mathcal{Q}_1({}^3S^{[1]}_1)\rangle_\Upsilon
+\frac{dF_4({}^3S^{[1]}_1)}{m^6}
\langle \mathcal{Q}_2({}^3S^{[1]}_1)\rangle_\Upsilon\nonumber\\
&&
+\frac{dF({}^3S^{[8]}_1)}{m^2}\langle \mathcal{O}({}^3S^{[8]}_1)\rangle_\Upsilon
+\sum_{J=0,1,2}\frac{dF({}^3P^{[8]}_J)}{m^4}
\langle \mathcal{O}({}^3P^{[8]}_J)\rangle_\Upsilon,
\end{eqnarray}
where $m$ indicates the mass of the bottom quark $b$,
$\mathcal{O}({}^{2S+1}L^{[c]}_J)$ indicates the NRQCD
operator with the spectroscopic state ${}^{2S+1}L^{[c]}_J$ with
the spin $S$, orbital angular momentum $L$, total angular momentum $J$, and
color quantum number $c$, and
$\langle \mathcal{O}({}^{2S+1}L^{[c]}_J)\rangle_H\equiv%
\langle H|\mathcal{O}({}^{2S+1}L^{[c]}_J)|H\rangle$ indicates the
NRQCD matrix element averaged over the spin states of $H$.  The
color quantum number $c=1$ and $8$ in the NRQCD operators denote
the color-singlet and the color-octet respectively. The NRQCD operators in
the factorization formula (\ref{factorization-formula}) are defined
by~\footnote{The dimensional regularization in quarkonium
calculations including definitions of operators was first given in
Refs.\cite{Braaten:1997dr,Braaten:1997cp}}
\begin{subequations}
\label{ope-def}
\begin{eqnarray}
\mathcal{O}({}^3S^{[1]}_1)&=&
\psi^\dagger\bm{\sigma}\chi\cdot
\chi^\dagger\bm{\sigma}\psi,
\\
\mathcal{P}({}^3S^{[1]}_1)&=&
\frac{1}{2}\bigg[
\psi^\dagger\bm{\sigma}\chi\cdot
\chi^\dagger
\big(-\tfrac{i}{2}\overleftrightarrow{\bm{D}}\big)^2\bm{\sigma}\psi
+
\psi^\dagger
\big(-\tfrac{i}{2}\overleftrightarrow{\bm{D}}\big)^2\bm{\sigma}\chi\cdot
\chi^\dagger\bm{\sigma}\psi
\bigg]
,
\\
\mathcal{O}({}^3S^{[8]}_1)&=&
\psi^\dagger \bm{\sigma}T^a\chi\cdot
\chi^\dagger \bm{\sigma}T^a\psi,
\\
\mathcal{O}({}^1S^{[8]}_0)&=&
\psi^\dagger T^a\chi\cdot
\chi^\dagger T^a\psi,
\\
\mathcal{O}({}^3P^{[8]}_0)&=&
\frac{1}{d-1}
\psi^\dagger \big(-\tfrac{i}{2}\overleftrightarrow{\bm{D}}
\cdot\bm{\sigma}\big)T^a\chi\cdot
\chi^\dagger \big(-\tfrac{i}{2}\overleftrightarrow{\bm{D}}
\cdot\bm{\sigma}\big)T^a\psi,
\\
\mathcal{O}({}^3P^{[8]}_1)&=&
\psi^\dagger \big(-\tfrac{i}{2}\overleftrightarrow{{D}}%
{}^{[i,}\sigma^{j]}\big)T^a\chi\cdot
\chi^\dagger \big(-\tfrac{i}{2}\overleftrightarrow{{D}}%
{}^{[i,}\sigma^{j]}\big)T^a\psi,
\\
\mathcal{O}({}^3P^{[8]}_2)&=&
\psi^\dagger \big(-\tfrac{i}{2}\overleftrightarrow{{D}}%
{}^{(i,}\sigma^{j)}\big)T^a\chi\cdot
\chi^\dagger \big(-\tfrac{i}{2}\overleftrightarrow{{D}}%
{}^{(i,}\sigma^{j)}\big)T^a\psi,
\\
\label{ope-def-q1}
\mathcal{Q}_1({}^3S^{[1]}_1)&=&
\psi^\dagger
\big(-\tfrac{i}{2}\overleftrightarrow{\bm{D}}\big)^2\bm{\sigma}\chi\cdot
\chi^\dagger
\big(-\tfrac{i}{2}\overleftrightarrow{\bm{D}}\big)^2\bm{\sigma}\psi,
\\
\label{ope-def-q2}
\mathcal{Q}_2({}^3S^{[1]}_1)&=&
\frac{1}{2}\bigg[
\psi^\dagger\bm{\sigma}\chi\cdot
\chi^\dagger
\big(-\tfrac{i}{2}\overleftrightarrow{\bm{D}}\big)^4\bm{\sigma}\psi+
\psi^\dagger
\big(-\tfrac{i}{2}\overleftrightarrow{\bm{D}}\big)^4\bm{\sigma}\chi\cdot
\chi^\dagger\bm{\sigma}\psi\bigg],
\end{eqnarray}
\end{subequations}
where $\psi$ and $\chi$ are Pauli spinor fields that annihilates the
bottom quark and creates the bottom antiquark, respectively,
$\sigma^i$ denotes the Pauli matrix, $\overleftrightarrow{\bm{D}}$
indicates the gauge-covariant derivative, and $d=4-2\epsilon$
represents the space-time dimensions. In (\ref{ope-def}),
$A^{[i,}B^{j]}$ and $A^{(i,}B^{j)}$ indicate the antisymmetric
tensor and the symmetric traceless tensor respectively, which are
defined via~\cite{Bodwin:2012xc}
\begin{subequations}
\label{ope-sym-antisym}
\begin{eqnarray}
A^{[i,}B^{j]}&\equiv& \frac{1}{2}(A^iB^j-A^jB^i),\\
A^{(i,}B^{j)}&\equiv& \frac{1}{2}(A^iB^j+A^jB^i)-\frac{1}{d-1}\delta^{ij}A^kB^k.
\end{eqnarray}
\end{subequations}
In (\ref{factorization-formula}), we omit the term associated with the matrix element of
the operator
\begin{eqnarray}
\frac{1}{2}\bigg[\psi^\dagger
\bm{\sigma}\chi\cdot
\chi^\dagger\big(\overleftrightarrow{\bm{D}}\cdot g{\bm E}+g{\bm E}\cdot\overleftrightarrow{\bm{D}}\big)\bm{\sigma}
\psi-\psi^\dagger
\big(\overleftrightarrow{\bm{D}}\cdot g{\bm E}+g{\bm E}\cdot\overleftrightarrow{\bm{D}}\big)
\bm{\sigma}\chi\cdot
\chi^\dagger\bm{\sigma}\psi\bigg],
\end{eqnarray}
which is shown to be dependent and be a linear combination of
the matrix elements of the operators (\ref{ope-def-q1}) and (\ref{ope-def-q2}) accurate up to
relative order $v^4$~\cite{Bodwin:2002hg}.

In order to get the decay rate, both
the NRQCD matrix elements and the
short-distance coefficients in (\ref{factorization-formula})
should be determined. The NRQCD matrix elements have been
studied through many nonperturbative approaches, such as lattice
QCD~\cite{Bodwin:1993wf},
nonrelativistic quark model~\cite{Eichten:1995},
and fitting the experimental data~\cite{Bodwin:2007fz, Guo:2011tz}.

On the other hand, based on the factorization, the short-distance
coefficients can be perturbatively determined through matching the
decay rates of the relevant processes at parton level
in full QCD to these in NRQCD. At the leading order in
$\alpha_s$, the coefficients $dF(^1S^{[8]}_0)$, $dF(^3S^{[8]}_1)$,
and $dF(^3P^{[8]}_J)$ can be determined through the processes
$b\bar{b}(^1S^{[8]}_0) \to c\bar{c}g$, $b\bar{b}(^3S^{[8]}_1) \to
c\bar{c}$, and $b\bar{b}(^3P^{[8]}_J) \to c\bar{c}g$ respectively.
The color-singlet short-distance coefficients can be determined
through the process $b\bar{b}(^3S^{[1]}_1) \to c\bar{c}gg$. Among
these short-distance coefficients, $F(^3S^{[8]}_1)$ has been
calculated up to the next-to-leading order in $\alpha_s$ in
Ref.~\cite{Zhang:2008pr}. $F_1(^3S^{[1]}_1)$ and $F_2(^3S^{[1]}_1)$
are obtained in
Refs.~\cite{Fritzsch:1978ey,Parkhomenko:1994xi,Kang:2007uv} and
Refs.~\cite{Chen:2011ph,Chen:2012zzg}, respectively.

As mentioned in Ref.~\cite{Bodwin:2002hg}, since there exists
the relation $\langle\mathcal{Q}_2({}^3S^{[1]}_1)\rangle_{\Upsilon}=
\langle\mathcal{Q}_1({}^3S^{[1]}_1)\rangle_{\Upsilon}(1+\mathcal{O}(v^2))$, to order $v^4$, we
can only determine the sum of the short-distance
coefficients:~\footnote{In Refs.~\cite{Brambilla:2006ph,Brambilla:2008zg,Ma:2002eva},
the authors may provide a potential approach to distinguish
$dF_3(^3S^{[1]}_1)$ and $dF_4(^3S^{[1]}_1)$ in (\ref{short-def-df}), nevertheless,
it is enough for us to determine $dF(^3S^{[1]}_1)$ in this work. }
\begin{equation}
\label{short-def-df}
dF(^3S^{[1]}_1)\equiv dF_3(^3S^{[1]}_1)+dF_4(^3S^{[1]}_1).
\end{equation}
As we will see, in order to get the IR-independent coefficient
$dF(^3S^{[1]}_1)$, we are required to calculate
$dF(^3S^{[8]}_1)$ and $dF(^3P^{[8]}_J)$, and make use of the color-octet mechanism
to cancel the IR divergences appearing in the decay rate of the process
$b\bar{b}(^3S^{[1]}_1) \to c\bar{c}gg$.

\section{kinematic definitions and phase-space decompositions}
\label{sec:definitions}

\subsection{Kinematic definitions}
\label{sec:sec:kinematics}
We assign $m_c$ to the mass of charm quark, and
assign $p_1$ and $p_2$ to the momenta of the incoming bottom quark $b$ and
antiquark $\bar{b}$. The momenta satisfy the on-shell relations:
$p_1^2=p_2^2=m^2$. $p_1$ and $p_2$ can be expressed as linear combinations
of their total momentum $P$ and half their relative momentum $p$:
\begin{equation}
p_1=P/2+p,\,\,\,\,\,\,\,\,\,\,\,\,
p_2=P/2-p.
\end{equation}

We assign $l_1, l_2$ to the momenta of the final charm pair. Therefore the
momentum of the virtual gluon yields to $Q=l_1+l_2$.
In addition, we take $k_1$ to the momentum of the gluon in the process
$b\bar{b}\to c\bar{c}g$, and take $k_1, k_2$ to the momenta of the two gluons
in the process $b\bar{b}\to c\bar{c}gg$.

To facilitate the evaluation,
 we introduce a set of dimensionless variables
\begin{equation}
x_1=\frac{2k_1\cdot P}{P^2}, \,\, x_2=\frac{2k_2\cdot P}{P^2}, \,\,
z=\frac{Q^2}{P^2},\,\, r=\frac{4m_c^2}{M^2},
\label{def-variables}
\end{equation}
where $M$ denotes the mass of the bottomonium. At the leading order in $v$,
there is $M=2m$.
All the Lorentz invariant kinematic quantities can
be expressed in terms of these new variables.

\subsection{Phase-space decompositions}
\label{sec:sec:phase-space} In this subsection, we present the
techniques for phase-space calculations of the relevant processes.
We decompose each phase-space integral into two parts, which
is proved to significantly simplify the calculations in the
following section.

\subsubsection{$b\bar{b}({}^3S^{[1]}_1)\to c\bar{c}gg$}

The process $b\bar{b}({}^3S^{[1]}_1)\to c\bar{c}gg$
involves four-body phase-space integral, which can be expressed as
\begin{eqnarray}
 \int \!\! d\phi_4&=&\!\!\! \int\!\!\! {d^{d-1}k_1\over (2\pi)^{d-1}2k^{0}_1} {d^{d-1}k_2\over
(2\pi)^{d-1}2k^{0}_2} {d^{d-1}l_1\over (2\pi)^{d-1}2l^{0}_1} {d^{d-1}l_2\over
(2\pi)^{d-1}2l^{0}_2} (2\pi)^d\delta^d(P-k_1-k_2-l_1-l_2).
\label{PT-4}
\end{eqnarray}
As the treatment in Ref.~\cite{Chen:2011ph}, we decompose the space-space integral into
\begin{equation}
 \int \!\! d\phi_4=\!\!\!  \int\!\!{dz\over 2\pi}
 \int \!\! d\phi_{3-1}\!\! \int \!\! d\phi_{2-1},
\label{PT-4-decom}
\end{equation}
where $\int \!\! d\phi_{3-1}$ and $\int \!\! d\phi_{2-1}$ are defined via
\begin{subequations}
\begin{eqnarray}
\int \!\! d\phi_{3-1}&=&\!\!\!\int\!\!{d^{d-1}Q\over
(2\pi)^{d-1}2Q^{0}}{d^{d-1}k_1\over (2\pi)^{d-1}2k^{0}_1}{d^{d-1}k_2\over
(2\pi)^{d-1}2k^{0}_2}(2\pi)^d\delta^d(P-k_1-k_2-Q),\\
\label{PT-4-3}
\int \!\! d\phi_{2-1}&=&\!\! P^2\!\! \int\!\! {d^{d-1}l_1\over (2\pi)^{d-1}2l^{0}_1}{d^{d-1}l_2\over
(2\pi)^{d-1}2l^{0}_2}(2\pi)^d\delta^d(Q-l_1-l_2).
\label{PT-4-2}
\end{eqnarray}
\end{subequations}

On the other hand, we can also separate the squared amplitude of
the process into two parts:
the charm part and the bottom part, i.e.,
\begin{equation}
\sum_{pol,col}|{\cal M}(^3S^{[1]}_1)|^2={L}^{(ab)\mu\nu}{H}^{(ab)}_{\mu\nu}(^3S^{[1]}_1),
\label{As-3S1-1}
\end{equation}
where polarizations and colors of the initial and final states are summed,
$a,b$ denote the color indices, and
the charm part ${L}^{(ab)\mu\nu}$ is given by
\begin{eqnarray}
{L}^{(ab)\mu\nu}=\frac{\delta^{ab}}{2}\frac{g_s^2}{Q^4}{\rm Tr}[(\not\!l_1+m_c)\gamma^\mu(\not\!l_2-m_c)\gamma^\nu],
\label{charm-part}
\end{eqnarray}
and the bottom part ${H}^{(ab)}_{\mu\nu}(^3S^{[1]}_1)$ accounts for the remainder.
According to the current conservation, we have
\begin{equation}
\int\!\!  d\phi_{2-1}{L}^{(ab)\mu\nu}
=\frac{\delta^{ab}}{2}\bigg(\!\!-g^{\mu\nu}+\frac{Q^\mu Q^\nu}{Q^2}\bigg)\times L,
\label{charm-L}
\end{equation}
where the Lorentz invariance $L$ is explicitly calculated
\begin{equation}
\label{charm-L-exp}
L= {\alpha_s\over 3 z}(2+\frac{r}{z}) \sqrt{1-{r\over z}}.
\end{equation}
As we will see, since
$L$ is a common factor in the whole calculations,
we get (\ref{charm-L-exp}) in 4-dimensions.
As a consequence, the decay rate yields
\begin{equation}
\Gamma(^3S^{[1]}_1)=\frac{1}{2}\int\!\! \frac{dz}{2\pi}L\int\!\! d\phi_{3-1} \frac{\delta^{ab}}{2}
{\widetilde H}^{(ab)}_{\mu\nu}(^3S^{[1]}_1)
\bigg(-g^{\mu\nu}+\frac{Q^\mu Q^\nu}{Q^2}\bigg),
\label{PT-4-rate}
\end{equation}
where a symmetry factor $\tfrac{1}{2}$ is included to account for the indistinguishability
of the two gluons.
The second term in the parenthesis of (\ref{PT-4-rate}) does not contribute,
due to the current conservation.

It is accustomed to reduce $\int\!\!
d\phi_{3-1}$ into the integration over two dimensionless variables:
\begin{eqnarray}
\label{PV-4-3-simp}
\int\!\!
d\phi_{3-1}=\frac{P^2}{128\pi^3}\frac{f_{\epsilon}^2e^{2\epsilon\gamma_E}}
{\Gamma(2-2\epsilon)}\int\!\!\! dx_1dx_2
\bigg\{(x_1+x_2-1+z)[(1-x_1)(1-x_2)-z]\bigg\}^{-\epsilon}\!\!\!\!\!\!.\phantom{x}
\end{eqnarray}
where $f_\epsilon\equiv(\tfrac{4\pi\mu^2}{M^2}e^{-\gamma_E})^{\epsilon}$.
Here $\mu$ in $f_\epsilon$
represents the dimensional-regularization scale, and $\gamma_E$ denotes the Euler constant.
The boundaries of $z, x_1$, and $x_2$ are readily inferred
\begin{eqnarray}
 1-x_1-z \le x_2 \le {1-x_1-z \over 1-x_1},\ \ \ 0\le x_1 \le 1-z, \ \ \  r\le z \le 1.
 \label{PV-4-3-bound}
\end{eqnarray}
It is convenient to make a further change of variables:
\begin{equation}
x_1=(1-z)x,\,\,\,\,\,\,\,\,\,
x_2=\frac{(1-z)(1-x)[1-(1-z)xy]}{1-(1-z)x}.
\end{equation}
This change of variables is particularly useful in extracting
the relativistic corrections.
Through this transformation, the boundaries of the new variables are
simplified to
\begin{equation}
0\le x\le 1, \,\,\,\,\,\,\,\,\,\,\,\,\, 0\le y\le 1.
\end{equation}
Finally, the decay rate reduces to
\begin{eqnarray}
\Gamma(^3S^{[1]}_1)=
\frac{f_{\epsilon}^2e^{2\epsilon\gamma_E}}{\Gamma(2-2\epsilon)}
\int\!\! dzdxdy \bigg[\frac{1-(1-z)x}{y(1-y)x^2(1-x)^2}\bigg]^\epsilon(1-z)^{-4\epsilon}
L\times H(^3S^{[1]}_1),
\label{PV-4-3-rate1}
\end{eqnarray}
where
\begin{eqnarray}
\label{H-3S1-1}
H(^3S^{[1]}_1)\equiv\frac{1}{2}\frac{P^2(1-z)^3}{(4\pi)^4}
\frac{x(1-x)}{1-(1-z)x}\frac{\delta^{ab}}{2}(-g^{\mu\nu}){\widetilde H}^{(ab)}_{\mu\nu}(^3S^{[1]}_1).
\end{eqnarray}

\subsubsection{$b\bar{b}(^3P^{[8]}_J)\to c\bar{c}g$}

Analogously, we separate the phase-space integral of this process into
\begin{eqnarray}
 \int \!\! d\phi_{3-2}=  \int\!\!{dz\over 2\pi}
 \int \!\! d\phi_{2-1}\!\!\! \int \!\! d\phi_{2-2},
\label{PT-3}
\end{eqnarray}
where $\int \!\! d\phi_{2-1}$ is given in (\ref{PT-4-2})
and $\int \!\! d\phi_{2-2}$ indicates
\begin{eqnarray}
\int \!\! d\phi_{2-2}&=&\int\!\!{d^{d-1}Q\over
(2\pi)^{d-1}2Q^{0}}{d^{d-1}k_1\over (2\pi)^{d-1}2k^{0}_1}(2\pi)^d\delta^d(P-k_1-Q).
\label{PT-3-2}
\end{eqnarray}
$\int \!\! d\phi_{2-2}$ can be straightforward
integrated out as
\begin{eqnarray}
\label{PT-3-2-simp}
\int \!\! d\phi_{2-2}
=\frac{(1-z)^{1-2\epsilon}}{8\pi}\frac{\Gamma(1-\epsilon)}
{\Gamma(2-2\epsilon)\Gamma(1+\epsilon)}f_\epsilon,
\end{eqnarray}
Similarly, we separate the squared amplitude into the charm part and the bottom part
${H}^{ab}_{\mu\nu}(^3P^{[8]}_J)$,
and obtain the decay rate
\begin{eqnarray}
\Gamma=
\frac{\Gamma(1-\epsilon)f_\epsilon e^{\epsilon\gamma_E}}
{\Gamma(2-2\epsilon)}
\int\!\! dz (1-z)^{-2\epsilon} L\times H(^3P^{[8]}_J),
\label{PV-3-rate}
\end{eqnarray}
where
\begin{eqnarray}
H(^3P^{[8]}_J)&\equiv&\frac{\delta^{ab}}{2}\frac{(1-z)
(-g^{\mu\nu}){H}^{ab}_{\mu\nu}(^3P^{[8]}_J)}{16\pi^2}.
\label{H-3PJ}
\end{eqnarray}

\subsubsection{$b\bar{b}(^3S^{[8]}_1)\to c\bar{c}$}

The phase-space integral of this process is quite simple.
Therefore we directly present the decay rate
\begin{eqnarray}
\Gamma(^3S^{[8]}_1)=
\int\!\! dz\delta(1-z) H(^3S^{[8]}_1)\times L,
\label{PT-2-rate}
\end{eqnarray}
where $H(^3S^{[8]}_1)$ is defined via
\begin{eqnarray}
H(^3S^{[8]}_1)\equiv\frac{\delta^{ab}}{2}
\frac{(-g^{\mu\nu}){H}^{ab}_{\mu\nu}(^3S^{[8]}_1)}{P^2}.
\label{H-3S1-8}
\end{eqnarray}

\section{Determining the short-distance coefficients}
\label{sec:matching}
In this section, we determine the differential short-distance coefficients of $\Upsilon$
include decay into a charm pair at relative order $v^4$.
$dF(^3S^{[8]}_1)$, $dF(^3P^{[8]}_J)$, and $dF(^3S^{[1]}_1)$ can be determined through calculating
the decay rates of the perturbative processes
$b{\bar b}(^3S^{[8]}_1)\to c\bar{c}$,
$b{\bar b}(^3P^{[8]}_J)\to c\bar{c}g$, and
$b{\bar b}(^3S^{[1]}_1)\to c\bar{c}gg$, respectively.

\subsection{$S$-wave color-octet}

We first determine the differential short-distance coefficient
$dF({}^3S^{[8]}_1)$ through matching the decay rate of
$b\bar{b}({^3S^{[8]}_1})\to c\bar{c}$ in full QCD to that in
 NRQCD. We also carry out the computations for the
renormalization of the operator ${\mathcal O}(^3S^{[8]}_1)$,
which will produce mixing with the operator ${\mathcal
O}(^3P^{[8]}_J)$ at the next-to-leading order in $\alpha_s$ and
with the operator ${\mathcal O}(^3S^{[1]}_1)$ at the
next-to-next-to-leading order in $\alpha_s$. Moreover we consider
the renormalization of the operators ${\mathcal
O}(^3P^{[8]}_J)$, which will induce mixing with the operator
${\mathcal O}(^3S^{[1]}_1)$. The renormalized operators will be
utilized to cancel the IR divergences  through the color-octet
mechanism when we determine the differential short-distance
coefficients $dF(^3P^{[8]}_J)$ and $dF(^3S^{[1]}_1)$.

\subsubsection{$dF({^3S^{[8]}_1})$}

The corresponding factorization formula for $b\bar{b}({^3S^{[8]}_1})\to c\bar{c}$
is expressed as
\begin{eqnarray}
\label{factorization-3s1-8}
\frac{d\Gamma(^3S^{[8]}_1)}{dz}&=&
\frac{dF({}^3S^{[8]}_1)}{m^2dz}\langle\mathcal{O}({}^3S^{[8]}_1)
\rangle_{b\bar{b}(^3S^{[8]}_1)}\,.
\end{eqnarray}
Through calculations, we obtain the expression of $H(^3S^{[8]}_1)$ defined in (\ref{H-3S1-8})
as
\begin{eqnarray}
\label{H-3S1-8-exp}
H(^3S^{[8]}_1)=\frac{\pi \alpha_s}{2m^2}.
\end{eqnarray}
Inserting (\ref{H-3S1-8-exp}) into (\ref{PT-2-rate}), we immediately
deduce the differential decay rate in full QCD. By making use
of $\langle\mathcal{O}({}^3S^{[8]}_1)
\rangle_{b\bar{b}(^3S^{[8]}_1)}=1$, where the quark pair state is
non-relativistically normalized, we readily get
\begin{eqnarray}
\label{short-3s1-1}
\frac{dF({}^3S^{[8]}_1)}{dz}&=&\delta(1-z)\frac{\pi\alpha_s}{2}\times L.
\end{eqnarray}

\subsubsection{$^3S^{[8]}_1\to {}^3P^{[8]}_J$}
\begin{figure}[tb]
\begin{center}
\includegraphics[height=2.0cm]{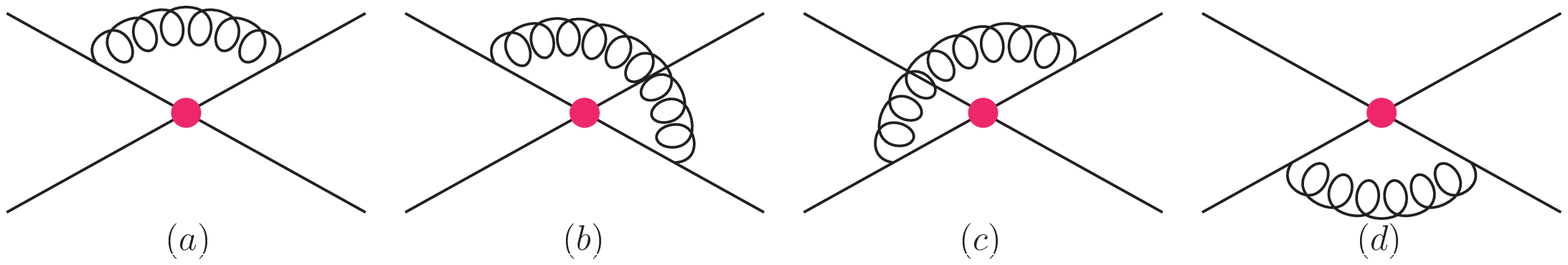}
\caption{The Feynman graphs for the NLO QCD corrections to the operator
${\mathcal O}(^3S^{[8]}_1)$. We suppress the graphs which do not give rise to
operator mixing.}
\label{fig-E1}
\end{center}
\end{figure}

In this subsection, we consider the NLO QCD corrections to the
operator ${\mathcal O}(^3S^{[8]}_1)$. There are four diagrams
illustrated in Fig.~\ref{fig-E1}~\footnote{Here we consider only the
diagrams which are relevant to our current work. Other diagrams will
take effect when one considers the NLO QCD corrections to the
short-distance coefficients. The similar calculations can also be
found in
Refs.~\cite{Petrelli:1997ge,Beneke:1998ks,Bodwin:2007zf,He:2009bf}}.
The vertex in the middle signifies the operator ${\mathcal
O}(^3S^{[8]}_1)$. Since the diagrams involve UV divergences, the
operator needs to be renormalized. In this work, we uniformly use
$\overline{\rm MS}$ scheme to carry out renormalization. We
express the renormalized operator as
\begin{eqnarray}\label{ope-ren-3s1}
{\mathcal O}(^3S^{[8]}_1)_{\overline{\rm MS}}={\mathcal O}(^3S^{[8]}_1)+\delta_{1} {}{\mathcal O}
+\delta_{2} {}{\mathcal O}+{\cal O}(\alpha_s^3),
\end{eqnarray}
where we truncate the expansion up to order $\alpha_s^2$, which is
enough to current work. $\delta_{1} {}{\mathcal O}$ and $\delta_{2}
{}{\mathcal O}$ are the corresponding NLO and the
next-to-next-to-leading-order (NNLO) counterterms respectively. For
convenience, we can expand an operator ${\mathcal O}$ in
terms of $\alpha_s$:
\begin{eqnarray}
\label{ope-expand}
{\mathcal O}={\mathcal O}^{(0)}+{\mathcal O}^{(1)}+
{\mathcal O}^{(2)}+\cdots.
\end{eqnarray}
where the superscript `(n)' represents order-$\alpha_s^{n}$
contribution.

In the following calculations, we will also utilize the color
decompositions~\cite{Petrelli:1997ge}
\begin{subequations}
\label{color-relation-1}
\begin{eqnarray}
T^aT^b\otimes T^aT^b &=&\frac{N_c^2-1}{4N_c^2}1 \otimes 1-\frac{1}{N_c}T^a \otimes T^a,\nonumber\\
T^aT^b\otimes T^bT^a &=&\frac{N_c^2-1}{4N_c^2}1 \otimes 1+\frac{N_c^2-2}{2N_c}T^a \otimes T^a.
\end{eqnarray}
\end{subequations}

With the NRQCD Feynman rules~\cite{Bodwin:1998mn},
the color-octet contribution from Fig.~\ref{fig-E1}(a) reads
\begin{eqnarray}
\label{I-a}
I_a&=&\frac{(ig_s)^2}{4m^2}\frac{-1}{N_c}T^a\otimes T^a\int \frac{d^dk}{(2\pi)^d}
\frac{i}{p^0-k^0-\frac{(\vec{{\bm p}}-\vec{{\bm k}})^2}{2m}+i\epsilon}
\frac{i}{p^{\prime0}-k^0-\frac{(\vec{{\bm p}^\prime}-\vec{{\bm k}})^2}{2m}+i\epsilon}\nonumber\\
&&\times\frac{4i({\bm p}\cdot{\bm p}^\prime-{\bm p}\cdot{\bm k}{\bm p}^\prime\cdot{\bm k}/{\bm k}^2)}
{k^2+i\epsilon}\nonumber\\
&=&-\frac{g_s^2}{2m^2}\frac{T^a\otimes T^a}{N_c}{\bm p}\cdot{\bm p}^\prime
\frac{d-2}{d-1}\int \frac{d^{d-1}k}{(2\pi)^{d-1}}\frac{1}{|{\bm k}|^3}\nonumber\\
&=&-\frac{\alpha_s}{3\pi m^2}\frac{T^a\otimes T^a}{N_c}\bigg(\frac{\tilde{f}_\epsilon}{\epsilon_{\rm UV}}-\frac{\tilde{f}_\epsilon}{\epsilon_{\rm IR}}\bigg){\bm p}\cdot{\bm p}^\prime,
\end{eqnarray}
where
$\tilde{f}_\epsilon\equiv(4\pi e^{-\gamma_E})^\epsilon$.
Since the coefficient is proportional to
$\frac{1}{\epsilon_{\rm UV}}-\frac{1}{\epsilon_{\rm IR}}$, the
factor $d$ coming from the loop integral in (\ref{I-a}) is replaced
with $4$ in the ${\overline{\rm MS}}$ scheme. Moreover, we add the
factor $\tilde{f}_\epsilon$ which is always associated with the
$\overline{\rm MS}$ scheme.

It is not hard to find that Fig.~\ref{fig-E1}(b-d) give the same contributions as Fig.~\ref{fig-E1}(a)
up to a color factor.
Summing all the contributions and employing (\ref{ope-ren-3s1}), we get
\begin{subequations}
\begin{eqnarray}
\label{ope-3s1-ren-1}
{\mathcal O}(^3S^{[8]}_1)_{\overline{\rm MS}}&=&{\mathcal O}(^3S^{[8]}_1)^{(0)}-
\frac{5\alpha_s}{9\pi m^2}\frac{\tilde{f}_\epsilon}{\epsilon_{\rm IR}}
\sum_J{\mathcal O}(^3P^{[8]}_J)+{\mathcal O}(\alpha_s^2),\\
\label{ope-3s1-counter-1}
\delta_{1}{\mathcal O}&=&-\frac{5\alpha_s}{9\pi m^2}\frac{\tilde{f}_\epsilon}{\epsilon_{\rm UV}}
\sum_J{\mathcal O}(^3P^{[8]}_J).
\end{eqnarray}
\end{subequations}

\subsubsection{${}^3P^{[8]}_J\to {}^3S^{[1]}_1$}
In the similar way, we get
\begin{eqnarray}
\label{ope-3pj-1}
{\mathcal O}(^3P^{[8]}_J)^{(1)}=
\frac{8\alpha_s}{27\pi m^2}\frac{{\bm p}^4N_J}{(d-1)^3}\bigg(\frac{\tilde{f}_\epsilon}{\epsilon_{\rm UV}}
-\frac{\tilde{f}_\epsilon}{\epsilon_{\rm IR}}\bigg)
{\mathcal O}(^3S^{[1]}_1)+{\mathcal O}(\alpha_s^2),
\end{eqnarray}
and
\begin{eqnarray}
\label{ope-3pj-ren-1}
{\mathcal O}(^3P^{[8]}_J)_{\overline{\rm MS}}={\mathcal O}(^3P^{[8]}_J)^{(0)}-
\frac{8\alpha_s}{27\pi m^2}\frac{{\bm p}^4N_J}{(d-1)^3}\frac{\tilde{f}_\epsilon}
{\epsilon_{\rm IR}}
{\mathcal O}(^3S^{[1]}_1)+{\mathcal O}(\alpha_s^2),
\end{eqnarray}
where $N_J$=$1$, $\frac{(d-1)(d-2)}{2}$, and $\frac{(d-2)(d+1)}{2}$ for $J=0, 1, 2$.
In (\ref{ope-3pj-1}) and (\ref{ope-3pj-ren-1}), we explicitly keep the factor $d$
stemming from the
$S$-wave extraction, i.e., ${p}^i{p}^j\to \frac{{\bm p}^2}{d-1}\delta^{ij}$.
The same treatment will be
implemented in the following section~\footnote{We thank G. T. Bodwin for
communications on this point.}.

\subsubsection{${}^3S^{[8]}_1\to {}^3S^{[1]}_1$}

\begin{figure}[tb]
\begin{center}
\includegraphics[height=8.0cm]{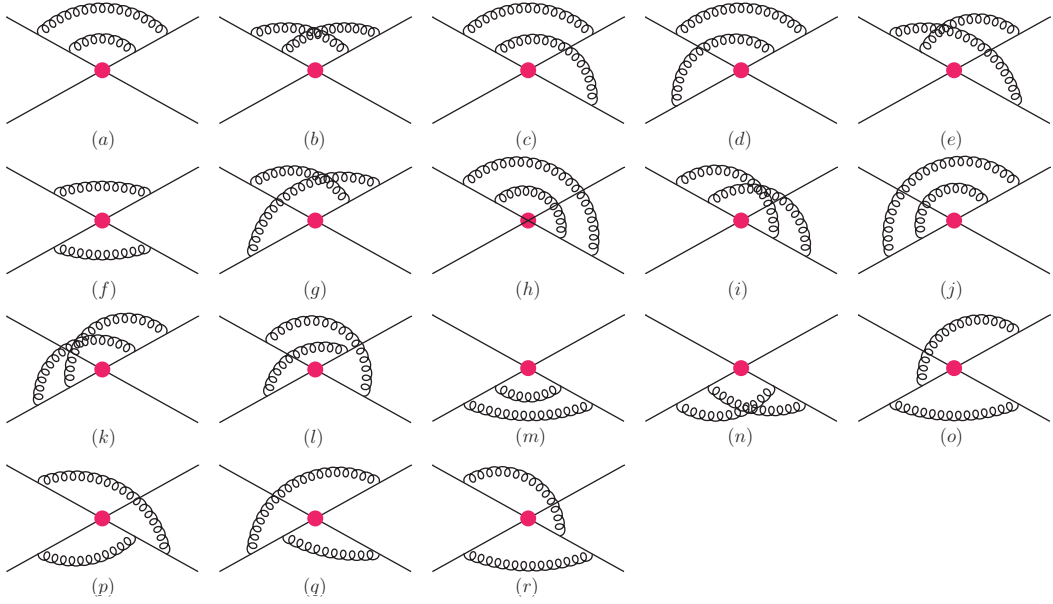}
\caption{The Feynman graphs for the NNLO QCD corrections to the
operator ${\mathcal O}(^3S^{[8]}_1)$.
We suppress the graphs which do not give rise to
operator mixing with ${\mathcal O}(^3S^{[1]}_1)$.}
\label{fig-2E1}
\end{center}
\end{figure}
We proceed to deal with the NNLO QCD corrections to the operator
${\mathcal O}(^3S^{[8]}_1)$. The relevant
Feynman graphs are illustrated
in Fig.~\ref{fig-2E1}. There are totally eighteen Feynman graphs which do contribution.
We are able to project out the color-singlet contribution by employing the color
decompositions
\begin{subequations}
\label{color-relation-2}
\begin{eqnarray}
T^aT^bT^c\otimes T^aT^bT^c &=&\frac{-2}{27}\, 1 \otimes 1+{\rm others},\\
T^aT^bT^c\otimes T^bT^aT^c &=&\frac{7}{27}\, 1 \otimes 1+{\rm others},
\end{eqnarray}
\end{subequations}
where we merely keep the color-singlet part. Through simple analysis, we find
the color factors are $\frac{7}{27}\, 1\otimes 1$ for the diagrams Fig.~\ref{fig-2E1}(a,e-g,i,k,m,p,q)
and $-\frac{2}{27}\, 1\otimes 1$ for the diagrams Fig.~\ref{fig-2E1}(b-d,h,j,l,n,o,r).

With the Feynman rules, the amplitude of Fig.~\ref{fig-2E1}(a)
reads
\begin{eqnarray}
\label{I-a-2}
I_{2a}&=&\frac{(ig_s)^4}{16m^4}\int\!\! \frac{d^dk}{(2\pi)^d}\int\!\! \frac{d^dl}{(2\pi)^d}
\frac{i}{p^0-k^0-\frac{(\vec{{\bm p}}-\vec{{\bm k}})^2}{2m}+i\epsilon}
\frac{i}{p^{\prime0}-k^0-\frac{(\vec{{\bm p}^\prime}-\vec{{\bm k}})^2}{2m}+i\epsilon}\nonumber\\
&&\times \frac{i}{p^0-k^0-l^0-\frac{(\vec{{\bm p}}-\vec{{\bm k}}-\vec{{\bm l}})^2}{2m}+i\epsilon}
\frac{i}{p^{\prime0}-k^0-l^0-\frac{(\vec{{\bm p}^\prime}-\vec{{\bm k}}-\vec{{\bm l}})^2}{2m}+i\epsilon}
\nonumber\\
&&\times\frac{4i({\bm p}\cdot{\bm p}^\prime-{\bm p}\cdot{\bm k}{\bm p}^\prime\cdot{\bm k}/{\bm k}^2)}
{k^2+i\epsilon}\frac{4i[({\bm p}-{\bm k})\cdot({\bm p}^\prime-{\bm k})
-({\bm p}-{\bm k})\cdot{\bm l}({\bm p}^\prime-{\bm k})\cdot{\bm l}/{\bm l}^2]}
{l^2+i\epsilon}\nonumber\\
&=&\frac{g_s^4}{4m^4}\frac{(d-2)^2}{(d-1)^2}\int\!\!
\frac{d^{d-1}k}{(2\pi)^{d-1}}\int\!\!\frac{d^{d-1}l}{(2\pi)^{d-1}}
\frac{({\bm p}\cdot{\bm p}^\prime)^2}
{|{\bm k}|^3|{\bm l}|(|{\bm k}|+|{\bm l}|)^2},
\end{eqnarray}
where the color factor is suppressed.

Analogous to (\ref{I-a-2}), we are able to get all others
\begin{eqnarray}
\label{I-bp-2}
I_{2b}&=&\frac{g_s^4}{4m^4}\frac{(d-2)^2}{(d-1)^2}\int\!\!
\frac{d^{d-1}k}{(2\pi)^{d-1}}\int\!\!\frac{d^{d-1}l}{(2\pi)^{d-1}}
\frac{({\bm p}\cdot{\bm p}^\prime)^2}
{|{\bm k}|^2|{\bm l}|^2(|{\bm k}|+|{\bm l}|)^2},\nonumber\\
I_{2c}&=&\frac{g_s^4}{4m^4}\frac{(d-2)^2}{(d-1)^2}\int\!\!
\frac{d^{d-1}k}{(2\pi)^{d-1}}\int\!\!\frac{d^{d-1}l}{(2\pi)^{d-1}}
\frac{({\bm p}\cdot{\bm p}^\prime)^2}
{|{\bm k}|^3|{\bm l}|^2(|{\bm k}|+|{\bm l}|)},\nonumber\\
I_f&=&\frac{g_s^4}{4m^4}\frac{(d-2)^2}{(d-1)^2}\int\!\!
\frac{d^{d-1}k}{(2\pi)^{d-1}}\int\!\!\frac{d^{d-1}l}{(2\pi)^{d-1}}
\frac{({\bm p}\cdot{\bm p}^\prime)^2}
{|{\bm k}|^3|{\bm l}|^3},\nonumber\\
I_h&=&I_j=I_m=I_a,\nonumber\\
I_i&=&I_k=I_n=I_b,\nonumber\\
I_d&=&I_e=I_g=I_o=I_p=I_q=I_r=I_c,\nonumber\\
I_l&=&I_f.
\end{eqnarray}

Including the color factors, and summing over all these
contributions, we get
\begin{eqnarray}
\label{ope-3s1-2}
I&=&\frac{5}{54}\frac{2g_s^4}{m^4}\frac{(d-2)^2}{(d-1)^2}\, 1 \otimes 1\int\!\!
\frac{d^{d-1}k}{(2\pi)^{d-1}}\int\!\!\frac{d^{d-1}l}{(2\pi)^{d-1}}
\frac{({\bm p}\cdot{\bm p}^\prime)^2}
{|{\bm k}|^3|{\bm l}|^3}\nonumber\\
&=&\frac{20\alpha_s^2}{243\pi^2(d-1)}\frac{{\bm p}^4}{m^4}\bigg(\frac{\tilde{f}_\epsilon}
{\epsilon_{\rm UV}}-\frac{\tilde{f}_\epsilon}{\epsilon_{\rm IR}}\bigg)^2 1 \otimes 1,
\end{eqnarray}
where we use ${{\bm p}^\prime}^2={\bm p}^2$.

In addition, we also need to calculate the NLO QCD corrections to the
counterterm $\delta_1{\mathcal O}$. Employing (\ref{ope-3pj-1}), we readily
obtain
\begin{eqnarray}
\label{ope-3s1-counter-1}
\delta_1{\mathcal O}^{(1)}&=&-\frac{5\alpha_s}{9\pi m^2}\frac{\tilde{f}_\epsilon}
{\epsilon_{\rm UV}}\sum_J{\mathcal O}(^3P^{[8]}_J)^{(1)}=
-\frac{40\alpha_s^2}{243\pi^2(d-1)}\frac{{\bm p}^4}{m^4}
\frac{\tilde{f}_\epsilon}{\epsilon_{\rm UV}}
\bigg(\frac{\tilde{f}_\epsilon}{\epsilon_{\rm UV}}
-\frac{\tilde{f}_\epsilon}{\epsilon_{\rm IR}}\bigg){\mathcal O}(^3S^{[1]}_1).\phantom{x}
\end{eqnarray}
Combining (\ref{ope-3s1-2}) and (\ref{ope-3s1-counter-1}), we get
\begin{eqnarray}
\label{ope-3s1-ren-2}
{\mathcal O}(^3S^{[8]}_1)^{(2)}&=&\frac{20\alpha_s^2}{243\pi^2}\frac{{\bm p}^4}{(d-1)m^4}\bigg(-\frac{\tilde{f}_\epsilon^2}{\epsilon_{\rm UV}^2}
+\frac{\tilde{f}_\epsilon^2}{\epsilon_{\rm IR}^2}\bigg){\mathcal O}(^3S^{[1]}_1)
+\delta_2{\mathcal O},
\end{eqnarray}
where the UV divergence can be canceled through the counterterm. Finally, we
present the renormalized operator and the corresponding counterterm
\begin{subequations}
\begin{eqnarray}
\label{ope-3s1-ren-2-msbar}
{\mathcal O}(^3S^{[8]}_1)_{\overline{\rm MS}}&=&{\mathcal O}(^3S^{[8]}_1)^{(0)}-
\frac{5\alpha_s}{9\pi m^2}\frac{\tilde{f}_\epsilon}
{\epsilon_{\rm IR}}
\sum_J{\mathcal O}(^3P^{[8]}_J)+
\frac{20\alpha_s^2}{243\pi^2}\frac{{\bm p}^4}{(d-1)m^4}
\frac{\tilde{f}_\epsilon^2}{\epsilon_{\rm IR}^2}{\mathcal O}(^3S^{[1]}_1),\\
\label{ope-3s1-3s1-delta2}
\delta_2{\mathcal O}&=&
\frac{20\alpha_s^2}{243\pi^2}\frac{{\bm p}^4}{(d-1)m^4}
\frac{\tilde{f}_\epsilon^2}{\epsilon_{\rm UV}^2}{\mathcal O}(^3S^{[1]}_1).
\end{eqnarray}
\end{subequations}
\subsection{$P$-wave color-octet}

In this subsection, we determine the short-distance coefficients $dF({}^3P^{[8]}_J)$ through
calculating the decay rates of the perturbative processes
$b\bar{b}(^3P^{[8]}_J)\to c\bar{c}g$. The factorization formulas for these processes are
expressed as
\begin{eqnarray}
\label{factorization-3pj}
d\Gamma(^3P^{[8]}_J)&=&\frac{dF(^3P^{[8]}_J)}{m^4}
\langle{\mathcal O}(^3P^{[8]}_J)\rangle_{b\bar{b}(^3P^{[8]}_J)}
+\frac{dF(^3S^{[8]}_1)}{m^2}\langle{\mathcal O}(^3S^{[8]}_1)\rangle_{b\bar{b}(^3P^{[8]}_J)}.
\end{eqnarray}

We can utilize (\ref{PV-3-rate}) to calculate the decay rates in
full QCD. First, we need to obtain $H(^3P^{[8]}_J)$ defined in
(\ref{H-3PJ}). To understand the IR structure and show the IR
cancelation, here we separate $H(^3P^{[8]}_J)$ into two parts:
$H(^3P^{[8]}_J)=H_d(^3P^{[8]}_J)+H_s(^3P^{[8]}_J)$.
$H_d(^3P^{[8]}_J)$ include the terms proportional to $1/(1-z)$,
which contribute the whole IR divergences to the decay rates
from the region with the real gluon being soft.
$H_s(^3P^{[8]}_J)$ take in charge of the remainder, which is absent
of any singularity. Correspondingly, we use the subscripts $d$, $s$
in both $d\Gamma(^3P^{[8]}_J)$ and $dF(^3P^{[8]}_J)$ to denote
the contributions from the two parts. The advantage of this
classification will be recognized when we determine the
short-distance coefficient $dF(^3S^{[1]}_1)$.

\subsubsection{$H_d(^3P_J^{[8]})$}

We directly present the expressions of $H_d(^3P^{[8]}_J)$
\begin{eqnarray}
\label{H-3PJ-d-exp}
H_d({}^3P^{[8]}_J)&=&\frac{5\alpha_s^2 (1- \epsilon){\bm p}^2}
{3m^4 (1-z) (3- 2 \epsilon )},
\end{eqnarray}
which are the same for $J=0, 1, 2$.

It is useful to make the following expansion
\begin{eqnarray}
\label{plus-f-4}
\frac{1}{(1-z)^{1+2\epsilon}}=-\frac{(1-r)^{-2\epsilon}}
{2\epsilon}\delta(1-z)+\bigg(\frac{1}{1-z}\bigg)_+
-2\epsilon\bigg(\frac{\ln(1-z)}{1-z}\bigg)_++{\mathcal O}(\epsilon^2),
\end{eqnarray}
where the plus functions are defined via
\begin{eqnarray}
\int_r^1\!\! dz (a)_+ f(z)\equiv\int_r^1\!\! dz a \bigg[f(z)-f(1)\bigg].
\label{plus-f-2}
\end{eqnarray}
Inserting (\ref{H-3PJ-d-exp}) into (\ref{PV-3-rate}) and
employing (\ref{plus-f-4}), we are able
to get the differential decay rates $d\Gamma_d({}^3P^{[8]}_J)$, which embrace
IR divergences.
Incorporating the factorization formulas (\ref{factorization-3pj}) and
the expressions (\ref{short-3s1-1})
and (\ref{ope-3s1-ren-1}), we find the IR divergences appearing on the
left hand side of (\ref{factorization-3pj}) are
exactly canceled by the renormalized $S$-wave color-octet matrix element on the right hand side.
It renders the short-distance coefficients $dF(^3P^{[8]}_J)$ free of
any singularity:
\begin{eqnarray}
\label{short-3pj-d}
\frac{dF_{d}(^3P^{[8]}_J)}{dz}&=&-\frac{5\alpha_s^2L f_\epsilon}{18}
\bigg\{\bigg[\frac{1}{\epsilon_{\rm IR} }\bigg(1-\frac{\tilde{f}_\epsilon}{f_\epsilon}\bigg)
+\frac{5}{3}-2 \ln (1-r)+\bigg(2 \ln
   ^2(1-r)-\frac{10}{3} \ln (1-r)\nonumber\\
   &&-\frac{\pi^2}{4}+\frac{28}{9}\bigg) \epsilon\bigg]\delta(1-z)-\bigg(\frac{1}{1-z}\bigg)_+
\bigg[2 +\frac{10 \epsilon }{3}\bigg]
+4\epsilon\bigg(\frac{\ln(1-z)}{1-z}\bigg)_+\bigg\}.
\end{eqnarray}
In (\ref{short-3pj-d}), we keep the terms linearly dependent on $\epsilon$, which will
induce finite contributions to the short-distance coefficient $dF(^3S^{[1]}_1)$.

\subsubsection{$H_s({}^3P^{[8]}_J)$}
We present the expressions of $H_s({}^3P^{[8]}_J)$ as
\begin{subequations}
\label{H-3PJ-s-exp}
\begin{eqnarray}
H_s({}^3P^{[8]}_0)&=&\frac{5\alpha_s^2 (5-z) (1-\epsilon ){\bm p}^2}{12m^4 (3-2 \epsilon)},\\
H_s({}^3P^{[8]}_1)&=&-\frac{5\alpha_s^2[2+z-2(1+z)\epsilon]{\bm p}^2}{6m^4 (3- 2 \epsilon)},\\
H_s({}^3P^{[8]}_2)&=&-\frac{5\alpha_s^2[4+z-(1+z)\epsilon]{\bm p}^2}{6m^4 (15
-16 \epsilon+4 \epsilon ^2)}.
\end{eqnarray}
\end{subequations}
The corresponding differential decay rates can be achieved by multiplying
a constant factor:
\begin{subequations}
\label{short-3pj-s}
\begin{eqnarray}
\frac{dF_{s}(^3P^{[8]}_0)}{dz}&=&\frac{5(5-z)\alpha_s^2L f_\epsilon}{36}
\bigg[1-2\epsilon \ln (1-z)+\frac{5\epsilon}{3}\bigg],\\
\frac{dF_{s}(^3P^{[8]}_1)}{dz}&=&-\frac{5\alpha_s^2L f_\epsilon}{18}\bigg[z+2-2\epsilon(z+2)\ln(1-z)
+\frac{2(z+5)}{3}\epsilon\bigg],\\
\frac{dF_{s}(^3P^{[8]}_2)}{dz}&=&-\frac{\alpha_s^2L f_\epsilon}{18}\bigg[z+4-2\epsilon(z+4)\ln(1-z)+
\frac{1}{15} (31 z+169)\epsilon\bigg].
\end{eqnarray}
\end{subequations}

\subsection{$S$-wave color-singlet}
In this section, we determine the short-distance coefficient
$dF({}^3S^{[1]}_1)$ through calculating the decay rate of the process
$b\bar{b}({}^3S^{[1]}_1)\to c\bar{c}gg$.
The corresponding factorization formula
is expressed as
\begin{eqnarray}
\label{factorization-3s1-1}
d\Gamma({}^3S^{[1]}_1)&=&
\frac{dF_1({}^3S^{[1]}_1)}{m^2}\langle \mathcal{O}({}^3S^{[1]}_1)\rangle_H
+\frac{dF_2({}^3S^{[1]}_1)}{m^4}\langle \mathcal{P}({}^3S^{[1]}_1)\rangle_H
\nonumber\\&&
+\frac{dF_{3}({}^3S^{[1]}_1)}{m^6}
\langle \mathcal{Q}_1({}^3S^{[1]}_1)\rangle_H
+\frac{dF_4({}^3S^{[1]}_1)}{m^6}
\langle \mathcal{Q}_2({}^3S^{[1]}_1)\rangle_H
\nonumber\\&&
+\frac{dF({}^3S^{[8]}_1)}{m^2}\langle \mathcal{O}({}^3S^{[8]}_1)\rangle_H
+\sum_J\frac{dF({}^3P^{[8]}_J)}{m^4}
\langle \mathcal{O}({}^3P^{[8]}_J)\rangle_H,
\end{eqnarray}
where the subscript `$H$' in the matrix elements represents $b\bar{b}({}^3S^{[1]}_1)$.
As mentioned in Sec.~\ref{sec:NRQCD-factorization},
the two matrix elements in the second line
of (\ref{factorization-3s1-1}) are equal at relative order $v^4$. Therefore, we
will determine the combined short-distance coefficient $dF({}^3S^{[1]}_1)$ defined
in (\ref{short-def-df})
in this subsection.

There are six
diagrams for this process.
The formula for the decay rate is given in (\ref{PV-4-3-rate1}).
Firstly, we need to calculate $H(^3S^{[1]}_1)$, which is defined
in (\ref{H-3S1-1}).
To separate the relativistic corrections~\footnote{Since in the region $z\to r$,
the decay rate will develop a logarithmic dependence on $r$, i.e., $\ln r$, which is
sensitive to the value of $r$,
we will not expand the $r$ appearing in $L$. For further explanations, we refer the reader to
Ref.~\cite{Chen:2011ph}.},
we expand $H(^3S^{[1]}_1)$
in powers of ${\bm p}^2$:
\begin{eqnarray}
H(^3S^{[1]}_1)=H^{(0)}(^3S^{[1]}_1)+H^{(2)}(^3S^{[1]}_1)\frac{{\bm p}^2}{m^2}+H^{(4)}(^3S^{[1]}_1)\frac{{\bm p}^4}{m^4}+{\cal O}({\bm p}^6),
\end{eqnarray}
where the first two orders have been considered in
Ref.~\cite{Chen:2011ph}. Our remainder task is to calculate
$H^{(4)}(^3S^{[1]}_1)$ and the corresponding decay rate. At order
$v^4$, the decay rate involves IR divergences. Analogous to the
treatment in the previous subsection, we separate
$H^{(4)}(^3S^{[1]}_1)$ into three parts:
$H^{(4)}(^3S^{[1]}_1)=H^{(4)}_d(^3S^{[1]}_1)+H^{(4)}_s(^3S^{[1]}_1)+H^{(4)}_r(^3S^{[1]}_1)$.
$H^{(4)}_d(^3S^{[1]}_1)$ proportional to $1/(1-z)$ contributes the
whole IR divergences to the decay rate in the region where the two
real gluons are simultaneous soft. In the following, we will
demonstrate that the IR divergences can be thoroughly canceled by
the renormalized $S$-wave color-octet matrix element
(\ref{ope-3s1-ren-2-msbar}) with the short-distance coefficient
(\ref{short-3s1-1}), together with the renormalized $P$-wave matrix
elements (\ref{ope-3pj-ren-1}) with the short-distance coefficients
$dF_d(^3P^{[8]}_J)$ given in (\ref{short-3pj-d}).
$H^{(4)}_s(^3S^{[1]}_1)$ contributes the whole IR divergence to the
decay rate in the region where only one of the real gluons is soft,
as a result, it should be proportional to either $1/x$ or $1/(1-x)$.
We will show that the IR divergence can be thoroughly canceled by
the renormalized $P$-wave matrix elements (\ref{ope-3pj-ren-1}) with
short-distance coefficients $dF_s(^3P^{[8]}_J)$ given in
(\ref{short-3pj-s}). $H^{(4)}_r(^3S^{[1]}_1)$ takes in charge of the
remainder and therefore corresponds to a finite contribution to the
decay rate. Similarly, we use the subscripts $d$, $s$, $r$ in
both $d\Gamma(^3S^{[1]}_1)$ and $dF(^3S^{[1]}_1)$ to denote the
contributions from the three parts.

\subsubsection{$H^{(4)}_{d}(^3S^{[1]}_1)$}
In this subsection,
we calculate the decay rate and short-distance coefficient related to $H^{(4)}_d(^3S^{[1]}_1)$.
First, we present the expression
\begin{eqnarray}
H^{(4)}_d(^3S^{[1]}_1)=\frac{40\alpha_s^3}{243 m^2 \pi
   x (1-x) (1-z)}\bigg[3(2
   y^2-2 y+1)+(8 y^2-8 y+1) \epsilon-8 y(1-y) \epsilon ^2 \bigg].\phantom{x}
\label{H-3S1-1-exp}
\end{eqnarray}
This term is proportional to $1/(1-z)$. In addition, we notice
it also contains the factor $1/x(1-x)$.
Therefore, we expect it will contribute a double IR pole to the decay rate
at the endpoints of $z$ and $x$.

Employing the expansion
\begin{eqnarray}
\frac{1}{(1-z)^{1+4\epsilon}}=-\frac{(1-r)^{-4\epsilon}}{4\epsilon}\delta(1-z)+\bigg(\frac{1}{1-z}\bigg)_+
-4\epsilon\bigg(\frac{\ln(1-z)}{1-z}\bigg)_++{\mathcal O}(\epsilon^2),
\label{plus-f-1}
\end{eqnarray}
and integrating out the variables $x, y$, we are able to obtain the differential decay rate
\begin{eqnarray}
\frac{d\Gamma_{d}({}^3S^{[1]}_1)}{dz}&=&\frac{20\alpha_s^3L f_\epsilon^2{\bm p}^4}{243\pi m^6}
\bigg\{\bigg[\frac{1}{\epsilon ^2}
+\frac{4- 4\ln (1-r)}{\epsilon }+8 \ln ^2(1-r)-16 \ln (1-r)-\frac{7\pi^2}{6}+\frac{35}{3}\bigg]\nonumber\\
&&
\times\delta(1-z)-\bigg(\frac{1}{1-z}\bigg)_+\bigg[\frac{4}{\epsilon }+ 2 \ln z+16\bigg]
+16\bigg(\frac{\ln(1-z)}{1-z}\bigg)_+\bigg\}.
\label{dG-d-1}
\end{eqnarray}

Inserting (\ref{dG-d-1}) into the factorization formula
(\ref{factorization-3s1-1}), we see the IR divergences on the left
hand side are exactly canceled by the $S$-wave color-octet
contribution, together with the renormalized $P$-wave
color-octet matrix elements with the short-distance coefficients
$dF_d(^3P^{[8]}_J)$ on the right hand side. Straightforwardly, we
get
\begin{eqnarray}
\frac{dF_d(^3S^{[1]}_1)}{dz}&=&\frac{10\alpha_s^3L}{729\pi}\bigg\{\bigg[
\ln^2\!\!\frac{\mu^2}{M^2}+\frac{10-12\ln(1-r)}{3}\ln\!\frac{\mu^2}{M^2}
+4 \ln ^2(1-r)-\frac{20}{3} \ln (1-r)+\frac{25}{9}\nonumber\\
&&-\frac{2 \pi
   ^2}{3}\bigg]
\delta(1-z)-\bigg(\frac{1}{1-z}\bigg)_+\bigg[4\ln\!\frac{\mu^2}{M^2}+2 \ln z+\frac{20}{3}\bigg]
+8\bigg(\frac{\ln(1-z)}{1-z}\bigg)_+\bigg\},
\label{short-3s1-d}
\end{eqnarray}
where $\mu$ indicates the factorization scale.

\subsubsection{$H^{(4)}_s(^3S^{[1]}_1)$}
The expression of $H^{(4)}_s(^3S^{[1]}_1)$ reads
\begin{eqnarray}
H^{(4)}_s(^3S^{[1]}_1)&=&\frac{20\alpha_s^3}{243\pi m^2}
\bigg\{\frac{-8 y^2+8 y-z-3}{x}+\frac{2 (z-5) (y^2-y)+z-1}{x}\epsilon
-\frac{1}{(1-y+yz)^4}\nonumber\\
&&\times\frac{1}{1-x}\bigg[12 (1-z)^3 y^6+12 (z-5) (z-1)^2 y^5-6 (z^3-11 z^2+31 z-21) y^4\nonumber\\
&&-24 (z^2-5 z+6)
   y^3+(z^4+z^3+9 z^2-27 z+96) y^2+2 (z^3-2 z^2-3 z-18) y\nonumber\\
   &&+z^2+3 z+6+\epsilon\bigg(16 (1-z)^3 y^6+16 (z-5) (z-1)^2 y^5-2(z-1)\nonumber\\
   &&
   \times(z^2-34z+81)y^4-8
   (z^2-14 z+21) y^3-(z^4-z^3-9 z^2-3 z-92) y^2\nonumber\\
&&-2 (3 z^2+7 z+12)
   y-z^2+z+2\bigg) \bigg]
\bigg\}.
\label{h-4-s}
\end{eqnarray}
This term is proportional to either $1/x$ or $1/(1-x)$, therefore, it contributes
single poles to the decay rate at the endpoints of $x$, i.e., $x\to 0$ and
$x\to 1$. After integrating out the variables $x, y$, we have
\begin{eqnarray}
\frac{d\Gamma_{s}(^3S^{[1]}_1)}{dz}&=&\frac{20\alpha_s^3L f_\epsilon^2{\bm p}^4}{729\pi m^6}
\bigg\{\frac{3 z+5}{\epsilon }
+\frac{1}{3 (1-z)^4}\bigg[21 z^5-43 z^4-101 z^3+291 z^2-248 z+80 \nonumber\\
&&+3(3
   z^5-4 z^4+4 z^3+12 z^2-9 z+6) \ln z
   \bigg]-4
(3 z+5) \ln(1-z)\bigg\}.
\label{dG-s-1}
\end{eqnarray}

It is not hard to find that the IR divergence in (\ref{dG-s-1}) is exactly canceled
by the renormalized $P$-wave color-octet matrix elements with the short-distance coefficients
$dF_s({}^3P^{[8]}_J)$, as we expected. We present the final result
\begin{eqnarray}
\frac{dF_{s}(^3S^{[1]}_1)}{dz}&=&\frac{10\alpha_s^3L}{2187\pi}
\bigg\{(3 z+5)\ln\!\frac{\mu^2}{M^2}+
\frac{1}{3 (1-z)^4}\bigg[6 z^5-16 z^4-59 z^3+153 z^2-131 z+47 \nonumber\\
&&+3(3
   z^5-4 z^4+4 z^3+12 z^2-9 z+6) \ln z
   \bigg]-2
(3 z+5) \ln(1-z)\bigg\}.
\label{short-3s1-s}
\end{eqnarray}

\subsubsection{$H^{(4)}_r(^3S_1^{[1]})$}
Finally, we deal with $H^{(4)}_r(^3S^{[1]}_1)$, which will produce a
finite contribution to the decay rate. Since the expression of
$H^{(4)}_r(^3S^{[1]}_1)$ is both tedious and cumbersome, here we
merely present the differential short-distance coefficient as
\begin{eqnarray}
\frac{dF_{r}({}^3S^{[1]}_1)}{dz}&=&\frac{\alpha_s^3L}{262440\pi(1-z)^4}
\bigg\{15 {\rm tan}^{-1}\!\bigg(\sqrt{\frac{1-z}{z}}\bigg)
\bigg[6(86 z^5-279 z^4-2548 z^3+6718 z^2\nonumber\\
&&-5694z+1609) {\rm tan}^{-1}\!\bigg(\sqrt{\frac{1-z}{z}}\bigg)+\sqrt{z(1-z)}
   (13908-38215z+33375z^2\nonumber\\
&&-7296z^3-476z^4)\bigg]
-6(299 z^5+22640 z^4
-27848 z^3+16208 z^2-2015 z+700)\nonumber\\
&&\times \ln z
-(1-z)(68916 z^4
-256073 z^3+696819 z^2
-669618 z+229580)
\bigg\}.\phantom{xxx}
\label{short-3s1-r}
\end{eqnarray}
\subsection{Summarizing the differential short-distance coefficients}
We summarize the differential short-distance coefficients calculated in the above:
\begin{subequations}\label{short-distance}
\begin{eqnarray}
\frac{dF({}^3S^{[8]}_1)}{dz}&=&\frac{\pi\alpha_s L}{2}\delta(1-z),\\
\frac{dF(^3P^{[8]}_J)}{dz}&=&\!\!\frac{-5\alpha_s^2L}{18}
\bigg\{\bigg[\ln\!\frac{\mu^2}{M^2}
+\frac{5}{3}-2 \ln (1-r)\bigg]\delta(1-z)-2\bigg(\frac{1}{1-z}\bigg)_++A_J
\bigg\},\\
\frac{dF(^3S^{[1]}_1)}{dz}&=&\frac{dF_d(^3S^{[1]}_1)}{dz}+\frac{dF_s(^3S^{[1]}_1)}{dz}
+\frac{dF_r(^3S^{[1]}_1)}{dz},
\end{eqnarray}
\end{subequations}
where $A_J$ are $-\tfrac{5-z}{2}$, $2+z$, and $\tfrac{4+z}{5}$ for $J=0,1,2$ respectively.

\section{Discussions and summaries\label{sec:discussion}}
\subsection{Discussions}
Applying the formulas (\ref{short-distance}) for the differential
short-distance coefficients obtained in the last section, we now
make some discussions.

\subsubsection{${dF(^3S^{[1]}_1)}/{dz}$ in the limit of $z\to 0$}
We first discuss the short-distance coefficient $dF(^3S^{[1]}_1)/dz$ in the limit of
$z\to 0$. It is not hard to derive
\begin{eqnarray}
\label{limit-z-0}
\frac{1}{L}\frac{dF(^3S^{[1]}_1)}{dz}\bigg|_{z\to 0}=
-\frac{35\alpha_s^3}{2187 \pi }\ln z-\frac{70\alpha_s^3}{2187 \pi } \ln\!\frac{\mu ^2}{M^2}+\frac{1609 \pi\alpha_s^3
   }{11664}-\frac{3913\alpha_s^3}{4374 \pi }.
\end{eqnarray}
In (\ref{limit-z-0}), we notice that the limitation bears the
logarithmic divergence $\ln z$. It is no surprise, owing to the fact
we actually do not regularize the singularity when $z$ approaches to
0. Moreover, we find that the coefficient of $\ln z$ equals exactly
to the corresponding coefficient of IR divergence in the decay rate
of the process $b\bar{b}(^3S^{[1]}_1)\to 3g$ up to a factor
$\frac{1}{2\pi}\times\frac{1}{2}$, as our expectation (The constant
factor originates in from (\ref{PT-4-rate}).). On the other hand,
when either of the two real gluons becomes soft, there exists IR
divergence which is regularized in dimensional regularization and
canceled by the renormalized $P$-wave color-octet matrix elements.
The term proportional to $\ln \!\frac{\mu ^2}{M^2}$ is related
to the divergence. We are delight with that the coefficient of
$\ln\!\frac{\mu ^2}{M^2}$ is exactly double that of $\ln z$.

\subsubsection{Color-singlet differential short-distance coefficients}

In (\ref{short-distance}), the two types of plus functions $(\tfrac{1}{1-z})_+$ and
$(\tfrac{\ln (1-z)}{1-z})_+$ diverge as $z\to 1$.
Since these singularities
arise when the momenta of the real gluons in the final states go to 0,
the distributions are actually unreliable in this region.
Nevertheless, the singularities in the
distributions are smeared when one integrates out $z$ and so the integrated
short-distance coefficients are well behaved.
In order to investigate
the dependence of the relativistic corrections on the virtuality of the intermediate gluon,
it is intriguing and enlightening to study two ratios:
$t_1(z)\equiv dF_2(^3S^{[1]}_1)/dF_1(^3S^{[1]}_1)$ and
$t_2(z)\equiv dF(^3S^{[1]}_1)/dF_1(^3S^{[1]}_1)$, where $dF_1(^3S^{[1]}_1)$ and $dF_2(^3S^{[1]}_1)$
are defined in (\ref{factorization-formula}) and
have been obtained in (53) of Ref.~\cite{Chen:2011ph}.
\begin{figure}[tb]
\begin{center}
\includegraphics[height=5.0cm]{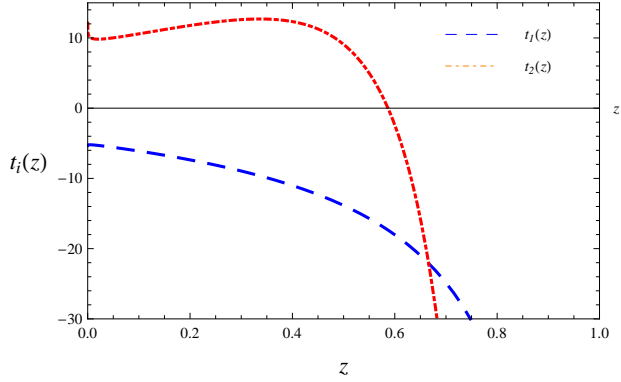}
\caption{Ratios of the differential short-distance coefficients.
The blue dashed line represents the
distribution $t_1(z)$, while the red dot-dashed line represents the distribution
$t_2(z)$. We specify the factorization scale $\mu=M$ in $t_2(z)$.}
\label{fig-distributions}
\end{center}
\end{figure}
To see it clearly, we plot Fig.~\ref{fig-distributions} to
illustrate the two ratios. In the figure, we observe that $t_1(z)$ which reflects the
NLO relativistic corrections is negative and
its magnitude rises quickly with increase of the virtuality of the immediate gluon.
However $t_2(z)$ which reflects the
order-$v^4$ relativistic corrections
is positive in small values of $z$ and turns to
negative in large values. The magnitude of $t_2(z)$ is sizable in most values of $z$.

\subsubsection{Integrated color-singlet short-distance coefficients}

\begin{figure}[tb]
\begin{center}
\includegraphics[height=5.0cm]{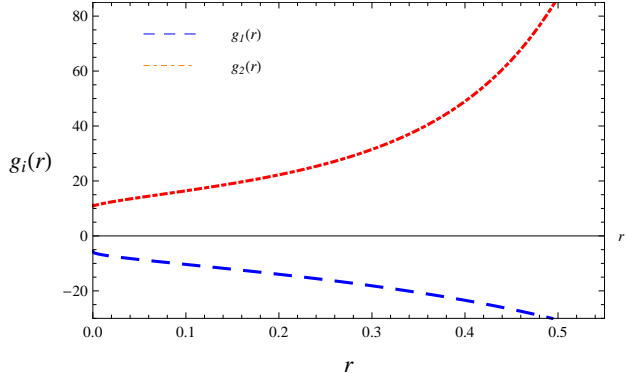}
\caption{Ratios of the short-distance coefficients as functions of $r$.
The blue dashed line represents the distribution $g_1(r)$, while the red dot-dashed line represents the distribution
$g_2(r)$. We specify the factorization scale $\mu=M$ in $g_2(r)$.}
\label{fig-gr}
\end{center}
\end{figure}

Finally, we integrate out the variable $z$ and investigate the integrated short-distance coefficients.
In Table.~{\ref{tab-raitos}}, we list the ratios of the order-$v^2$ and the order-$v^4$ color-singlet
short-distance coefficients to
the LO one for the processes of $\Upsilon(nS)$ inclusive decay into a charm pair.
\begin{table}
\centering
\caption{\label{tab-raitos} Ratios of the short-distance coefficients for
$\Upsilon\to c\bar{c}gg$. The charm mass is selected to that of the $D$
meson~\cite{Kang:2007uv,Chen:2011ph}.
The masses of the $D$ meson and bottomonia are taken from Ref.~\cite{Nakamura:2010zzi}.}
\begin{tabular}{lccccccccc}
\hline
&  $r$ & $F_2(^3S^{[1]}_1)/F_1(^3S^{[1]}_1)$ &$F(^3S^{[1]}_1)/F_1(^3S^{[1]}_1)$\\
\hline
$\Upsilon (1S)$ & $1.56\times10^{-1}$ & $-12.4$ & $19.4+0.6\ln^2(\frac{\mu^2}{M^2})
+0.9\ln(\frac{\mu^2}{M^2})$\\
\hline
$\Upsilon (2S)$ &$1.39\times10^{-1}$ &$-11.7$ & $18.5+0.5\ln^2(\frac{\mu^2}{M^2})
+0.6\ln(\frac{\mu^2}{M^2})$\\
\hline
$\Upsilon (3S)$ &$1.30\times10^{-1}$ &$-11.4$ & $18.0+0.5\ln^2(\frac{\mu^2}{M^2})
+0.4\ln(\frac{\mu^2}{M^2})$\\
\hline
\end{tabular}
\end{table}
We learn from the table that the ratio
$F(^3S^{[1]}_1)/F_1(^3S^{[1]}_1)$ is both positive and sizable,
nevertheless, the relativistic expansion is of well convergence
due to a small value of $v$ (e.g., $v^2\sim 0.1$.). The
situation is quite similar to the case for the process $J/\psi\to
ggg$~\cite{Bodwin:2002hg}. To further study the relation between the
relativistic corrections and $r$, we extrapolate the value of $r$
and investigate the ratios: $g_1(r)\equiv
F_2(^3S^{[1]}_1)(r)/F_1(^3S^{[1]}_1)(r)$ and $g_2(r)\equiv
F(^3S^{[1]}_1)(r)/F_1(^3S^{[1]}_1)(r)$. We illustrate the two
functions in Fig.~\ref{fig-gr}.
From the figure, we find the NNLO relativistic corrections become
more and more important with $r$ increase.

\subsection{Summaries}

In this work, we determine the short-distance coefficients within
the framework of NRQCD factorization formula for $\Upsilon$
inclusive decay into a charm pair through relative order $v^4$. The
order-$v^4$ color-singlet differential short-distance coefficient
$dF(^3S^{[1]}_1)$ is obtained through matching the decay rate of
$b\bar{b}(^3S^{[1]}_1)\to c\bar{c}gg$ in full QCD to that in
NRQCD. The double and single IR divergences appearing in the
decay rate are exactly canceled through the NNLO renormalization
of the operator ${\mathcal O}(^3S^{[8]}_1)$ and the NLO
renormalization of the operators ${\mathcal O}(^3P^{[8]}_J)$. To
investigate the magnitude of the relativistic corrections and the
convergence of the relativistic expansion, we show both the
ratios of the differential short-distance coefficients $t_i(z)$ and
the ratios of the short-distance coefficients $g_i(r)$. Our results
indicate that though $F(^3S^{[1]}_1)$ is quite large, the
relativistic expansion from the color-singlet contributions in
the process $\Upsilon\to c\bar{c}+X$ are well convergent due to
a small value of $v$. In addition, we extrapolate $g_i(r)$
to a large range of $r$, and find the relativistic corrections rise
quickly with increase of $r$.

\begin{acknowledgments}
We thank G. T. Bodwin for helpful discussions.
The research of W. S. was supported by China Postdoctoral Science Foundation
under Grant No. 2012M510549.
The research of H. C. and Y. C. was supported by the NSFC with Contract No. 10875156.
\end{acknowledgments}


\end{document}